\documentclass[usenatbib]{mn2e_add}
\usepackage{graphicx}
\usepackage{natbib}
\usepackage{color}
\bibpunct{(}{)}{;}{a}{}{,}
\usepackage{txfonts}

\newcommand{\dd}{$\rm deg^{2}$}
\newcommand{\flux}{$\rm erg \, s^{-1} \, cm^{-2}$}

\begin{document}

\title[The XMM-LSS D1 cluster sample]
{The XMM-LSS Survey: A well controlled X-ray cluster sample over
the D1 CFHTLS area \thanks{Based on data collected with XMM, VLT,
Magellan, NTT, and CFH telescopes; ESO programme numbers are:
070.A-0283, 070.A-907 (VVDS), 072.A-0104, 072.A-0312, 074.A-0360,
074.A-0476 }}
\author[M. Pierre et al]{M. Pierre$^{1}$\thanks{E-mail:
M. Pierre@cea.fr },  F. Pacaud$^{1}$, P.-A. Duc$^{1}$, J.P.
Willis$^{2}$, S. Andreon$^{3}$, I. Valtchanov$^{4}$\thanks{current
address: ESA, Villafranca del Castillo, Spain}
\newauthor
  B. Altieri$^{5}$, G. Galaz$^{6}$,  A. Gueguen$^{1}$, J.-P. Le
F\`evre$^{7}$, O. Le
F\`evre$^{8}$, T. Ponman$^{9}$,
\newauthor
  P.-G. Sprimont$^{10}$, J. Surdej$^{10}$,
\newauthor
  C. Adami$^{8}$, A. Alshino$^{9}$,  M. Bremer$^{11}$, L.
Chiappetti$^{12}$, A. Detal$^{10}$,
   O. Garcet$^{10}$, E. Gosset$^{10}$,
\newauthor
  C. Jean$^{10}$, D. Maccagni$^{12}$, C. Marinoni$^{8}$, A.  Mazure$^{8}$,   H.
Quintana$^{6}$, A. Read$^{13}$ \\
$^{1}$DAPNIA/SAp CEA Saclay, 91191 Gif sur Yvette\\
$^{2}$Department of Physics and Astronomy, University of Victoria,
Elliot Building, 3800 Finnerty Road, Victoria, BC, V8P 1A1
Canada.\\
$^{3}$INAF -- Osservatorio Astronomico di Brera,
Milan, Italy.\\
$^{4}$Astrophysics Group, Blackett Laboratory, Imperial College
of Science Technology and Medicine, London SW7 2BW, UK.\\
$^{5}$ESA, Villafranca del Castillo, Spain.\\
$^{6}$Departamento de Astronom{\'i}a y Astrof{\'i}sica, Pontificia
Universidad Cat{\'o}lica de Chile,
Santiago, Chile.\\
$^{7}$DAPNIA/SEDI CEA Saclay, 91191 Gif sur Yvette\\
$^{8}$ Laboratoire d'Astrophysique de Marseille, France\\
$^{9}$School of Physics and Astronomy,
University of Birmingham, Edgebaston, Birmingham B15 2TT, UK.\\
$^{10}$Universit{\'e} de Li{\`e}ge, All{\'e}e du 6 Ao{\^u}t, 17,
B5C, 4000 Sart Tilman, Li\`ege Belgium. \\
$^{11}$Department of Physics,
University of Bristol, Tyndall Avenue, Bristol BS8 1TL, UK.\\
$^{12}$INAF --  IASF Milan, Italy.\\
$^{13}$Department of Physics and Astronomy, University of
Leicester, Leicester LE1 7RH, UK.  }

\date{Accepted for publication in MNRAS - July 2006 }

\pagerange{\pageref{firstpage}--\pageref{lastpage}} \pubyear{2002}

\maketitle

\label{firstpage}

\begin{abstract}

We present the XMM-LSS cluster catalogue corresponding to the
CFHTLS D1 area. The list contains 13 spectroscopically confirmed,
X-ray selected galaxy clusters over 0.8 \dd\ to a redshift of
unity and so constitutes the highest density sample of clusters to
date.  Cluster X-ray bolometric luminosities range from 0.03 to $5
\times 10^{44}$ erg s$^{-1}$. In this study, we describe our
catalogue construction procedure: from the detection of X-ray
cluster candidates to the compilation of a spectroscopically
confirmed cluster sample with an explicit selection function. The
procedure further provides basic X-ray products such as cluster
temperature, flux and luminosity. We detected slightly more
clusters with a (0.5-2.0 keV) X-ray fluxes of $>2 \times 10^{-14}$
erg s$^{-1}$ cm$^{-2}$ than we expected based on expectations from
deep ROSAT surveys. We also present the Luminosity-Temperature
relation for our 9 brightest objects possessing a reliable
temperature determination. The slope is in good agreement with the
local relation, yet  compatible with a luminosity enhancement for
the $0.15 < z< 0.35$ objects having $1 < T <2 $ keV, a population
that the XMM-LSS is identifying systematically for the first time.
The present study permits the compilation of cluster samples from
XMM images whose selection biases are understood. This allows, in
addition to studies of large-scale structure, the systematic
investigation of cluster scaling law evolution, especially for low
mass X-ray groups which constitute the bulk of our observed
cluster population. All cluster ancillary data (images, profiles,
spectra) are made available in electronic form via the XMM-LSS
cluster database.

\end{abstract}


\section{Introduction}

The question of cosmic structure formation is substantially more
complicated than the study of the spherical collapse of a pure dark
matter perturbation in an expanding Universe.  While it is possible to
predict theoretically how the shape of the inflationary fluctuation
spectrum evolves until recombination, understanding the subsequent
formation of galaxies, AGN and galaxy clusters is complicated by the
physics of non-linear growth and feedback from star
formation. Attempts to use the statistics of visible matter
fluctuations to constrain the nature of Dark Matter and Dark Energy
are therefore reliant upon an understanding of non-gravitational
processes.

Clusters, as the most massive entities of the Universe, form a crucial
link in the chain of understanding. They lie at the nodes of the
cosmic network, possess virialized cores, yet are still growing by
accretion along filaments. The rate at which clusters form, and the
evolution of their space distribution, depends strongly on the shape
and normalization of the initial power spectrum, as well as on the
Dark Energy equation of state (e.g. \citealt{rapetti05}).  Consequently,
both a three dimensional map of the cluster distribution and an
evolutionary model relating cluster observables to cluster masses and
shapes (predicted by theory for the average cluster population) are
needed to test the consistency of structure formation models within a
standard cosmology with the properties of clusters in the low-$z$
Universe.

The main goal of the XMM Large Scale Structure Survey (XMM-LSS) is
to provide a  well defined statistical  sample of X-ray galaxy
clusters to a redshift of unity over a single large area, suitable
for cosmological studies \citep{pierre04}. In this paper, we
present the first sample of XMM-LSS clusters for which canonical
selection criteria are uniformly applied over the survey area. In
this way, we demonstrate the properties of the survey along with a
description of data analysis tools employed in the sample
construction; the aim being to provide a deep and well controlled
sample of clusters and to investigate evolution trends, in
particular for the low end of the cluster mass function. The paper
will therefore act as a reference for future studies using XMM-LSS
data. The chosen region is located at $36 < {\rm R.A. (deg.)} < 37,
\,\, -5 < {\rm Dec. (deg.)} < -4 $. This region is known as D1, one of
the four Deep areas of the Canada-France-Hawaii-Telescope Legacy
Survey\footnote{http://cdsweb.u-strasbg.fr:2001/Science/CFHLS/}
(CFHTLS). It also includes one of the VIMOS VLT Deep Survey
patches (VVDS; \citealt{vvds}) and was observed at 1.4 GHz down to
the $\mu$Jy level by the VLA-VIRMOS Deep Field \citep{bondi03}.
The rest of the XMM-LSS survey surrounds D1 and corresponds to
part of the wide W1 CFHTLS component (see \citet{pierre04} for a
general lay-out and associated multi-$\lambda$ surveys) for which
the complete cluster catalogue will be published separately. The
sample is the result of a fine tuned X-ray plus optical approach
developed with the aim of understanding the various selection
effects. We describe the catalogue construction procedure in
tandem with a companion paper presenting a detailed description of
the X-ray pipeline developed as part of the XMM-LSS survey
\citep{pacaud06}.

The deepest published statistical  samples of X-ray clusters over
a contiguous sky area to date are all based on the ROSAT All-Sky
Survey (RASS): REFLEX \citep{bohringer01}, NORAS
\citep{bohringer00}, NEP \citep{henry01}. In parallel, a number of
serendipitous cluster surveys were conducted using deep ROSAT
pointings with the goal of investigating the evolution of the
cluster luminosity function e.g. Southern SHARC \citep{burke97},
RDCS \citep{rosati98}, 160 \dd\ \citep{vikhlinin98}, Bright SHARC
\citep{romer00}, BMW \citep{moretti04}.  The advent of the XMM
satellite has provided an X-ray imaging capability of increased
sensitivity and angular resolving power compared to ROSAT. The
XMM-LSS employs 10-20 ks pointings and samples the cluster
population to a depth of $\sim 10^{-14}$ \flux\ \---\ a flux
sensitivity comparable to the deepest serendipitous ROSAT surveys
\citep{rosati02}. However, XMM observations possess a narrower PSF
(FWHM $\sim6\arcsec$ for XMM vs $\sim 20\arcsec$ for the ROSAT
PSPC) which suggests that the reliable identification of extended
sources can be performed for apparently smaller sources.
Instrumental characteristics such as background noise and the
complex focal plane configuration are also quite different. In
this context, our dual aim of optimizing the XMM-LSS sensitivity
and of quantifying the many selection biases led us to develop a
dedicated source detection pipeline as well as specific optical
identification and spectroscopic confirmation procedures: special
attention is given to extended, X-ray faint sources whose
identification requires deep optical/IR multi-color imaging. These
steps are described in Section 2 along with the presentation of
the D1 cluster catalogue. Section 3 presents the X-ray properties
of the newly assembled sample and some optical characteristics.
Section 4 summarizes the global properties of our sample within
the context of a ``concordance'' cosmological model. We conclude
with a discussion of our cluster selection function in comparison
with earlier works as well as the scaling laws for the low end of
the cluster mass function.

Throughout the paper we assume $\Omega_{M}=0.27$,
$\Omega_{\Lambda}=0.73$ and $H_0 = 71$~km s$^{-1}$ Mpc$^{-1}$. All
X-ray flux measures are quoted in the [0.5-2] keV band. The generic
name ``cluster'' refers to the entire population of gravitationally
bound galaxy systems, while we use the term ``groups'' for those
systems whose potential corresponds to an X-ray temperature lower than
2 keV.


\section{The X-ray cluster catalogue }

\subsection{X-ray observations}

\begin{figure}
   \centering
   \includegraphics[width=8cm]{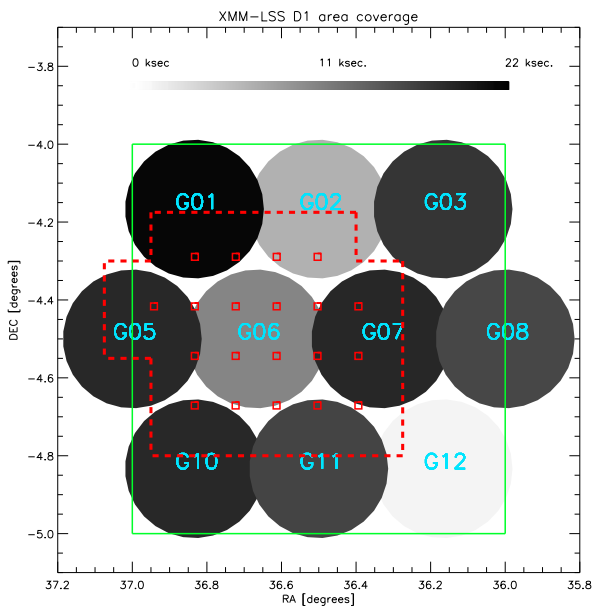}
      \caption{The XMM pointing mosaic over the D1 area (green
      square).  The radius of the displayed pointings is 11
      arcminutes.  The grey-scale indicates effective mean exposure
      time per detector, after removal of high background periods.
      The red squares show the centres of the VVDS pointings
      \citep{vvds} and the red dotted line indicates the total area
      covered covered by the VVDS. The VLA-VIRMOS Deep Field
      encompasses exactly the D1 region}
         \label{mosaic}
   \end{figure}

The XMM-LSS D1 region consists of a mosaic of 10 XMM pointings
that form part of the XMM Medium Deep Survey (XMDS;
\citealt{chiappetti05}). The pointing layout is displayed in
Figure \ref{mosaic} and the properties of individual pointings are
shown in Table \ref{pointing}. The nominal exposure per pointing
is 20 ks for this subregion\footnote{Outside the XMDS, the nominal
exposure per pointing for the rest of the XMM-LSS is 10 ks.}, with
the exception of pointing G07, whose nominal exposure time of 40
ks was reduced to $\sim 20$ effective ks as a result of solar
activity. The raw X-ray observations (ODFs) were reduced using the
standard XMM Science Analysis System (XMMSAS; version v6.1) tasks
{\tt emchain} and {\tt epchain} for the MOS and PN detectors
respectively. High background periods, related to soft protons,
were excluded from the event lists following the procedure
outlined by \citet{Pratt02}. Raw photon images in different energy
bands were then created with a scale of 2\farcs5 pixel$^{-1}$. A
complete discussion of the image analysis and source
characterization procedures are provided by \citet{pacaud06}.
Cluster detection was performed in the [0.5-2] keV band and was
limited to the inner 11 arcmin of the XMM field. The total scanned
area is 0.81 \dd. Information regarding the individual pointings
is summarized in Table \ref{pointing} and the layout of the
pointings on the sky displayed in Fig. \ref{mosaic}.

\begin{table*}
\caption{Properties of individual XMM pointings. Quoted exposures
are effective exposures computed after filtering high background
periods.}
 \label{pointing} \centering
\begin{tabular}{l l  l l l }
\hline
Internal ID & XMM ID & RA (J2000) & Dec. (J2000) & MOS1, MOS2, pn exposure times (ks)   \\
\hline\hline
    G01 & 0112680201  & 02:27:20.0 & -04:10:00.0 & 24.6, 25.3,
    21.4\\
    G02 & 0112680201  & 02:26:00.0 & -04:10:00.0 & 10.1, 9.7, 6.7
    \\
    G03 & 0112680301  & 02:24:40.0 & -04:10:00.0 & 21.8, 21.7,
    17.3 \\
    G05 & 0112680401  & 02:28:00.0 & -04:30:00.0 & 23.5,
    23.9, 12.5\\
    G06 & 0112681301  & 02:26:40.0 & -04:30:00.0 &16.4, 16.6,
    10.5 \\
    G07 &  0112681001 & 02:25:20.0 & -04:30:00.0 & 22.5, 25.1,
    18.6 \\
    G08 & 0112680501  & 02:24:00.0 & -04:30:00.0 & 21.2, 21.3,
    15.9 \\
    G10 & 0109520201  & 02:27:20.0 & -04:50:00.0 & 24.7, 24.6,
    18.5 \\
    G11 & 0109520301  & 02:26:00.0 & -04:50:00.0 & 21.7, 21.8,
    16.1 \\
    G12 & 0109520401  & 02:24:40.0 & -04:50:00.0  & not usable
    because of very high flare rate \\
\hline
\end{tabular}
\end{table*}

\subsection{Cluster X-ray detection and optical identification procedure}

The compilation of an X-ray cluster sample featuring positional
and redshift data ultimately requires the input of optical and/or
near-infrared (NIR) data in order to select putative cluster
galaxies for which precise redshifts can be obtained. Therefore,
although X-ray selection is employed to better avoid projection
effects, to provide direct clues about cluster masses and to
provide more easily tractable selection criteria, optical/NIR data
for each cluster must be assessed in order to obtain cluster
redshift data. The goal of XMM-LSS is to produce a faint,
statistical cluster catalogue over a wide spatial area (several
tens of square degrees) and a large redshift interval (zero to
unity). Cluster identification procedures must therefore identify
robustly a wide range of cluster properties at both X-ray and
optical/NIR wavelengths. Given the above requirements the XMM-LSS
has developed over the last three years from initially simple and
very robust cluster selection procedures to a refined,
quantitative approach focusing on key cluster selection
parameters.

Developing the X-ray pipeline was an essential part of the
procedure as is reflected in the successive publications. We
summarise these developments below:
\begin{enumerate}
\item Spectroscopic observations during 2002 were performed for a
number of cluster candidates identified following the method developed
by \citep{val01}; extended X-ray sources were accepted as candidate
clusters if associated with a spatial overdensity of galaxies
displaying a uniform red colour sequence determined using either
CFHT/CFH12K $BVRI$ or CTIO/MOSAIC $Rz$ imaging. This approach
maximized the success rate of the first XMM-LSS spectroscopic
observations (conducted in the last quarter of 2002) which
demonstrated that clusters to a redshift of 1 are detectable with
10-20 ks XMM observations (Valtchanov et al. 2004; Willis et
al. 2005).
\item In order to proceed toward a purely X-ray selected sample \---\
i.e. to reduce contamination by spurious extended sources \---\ a
maximum likelihood procedure named {\tt Xamin} was combined with the
wavelet-based detection algorithm developed previously
\citep{pierre04}. The sample of candidate clusters thus generated was
investigated during the spectroscopic observations conducted in 2003 and 2004.
\item Finally, the combination of spectroscopic results for the
above cluster sample with a detailed study of the simulated
performance of the X-ray pipeline led to the definition of three
clearly defined classes of X-ray cluster candidates.
\end{enumerate}

The cluster identification procedure described above satisfies the
goal of generating a relatively uncontaminated sample of X-ray
clusters with a well defined selection function.  A detailed description
of the X-ray parameters employed to generate each class of cluster
candidate is described in the following section.

\subsection{The cluster classification and sample}
\label{class}

The ability to detect faint, extended sources in X--ray images is
subject to  a number of factors. Although the apparent size of a
typical cluster ($R_{c} =180$~kpc) is significantly larger than
the XMM point spread function (PSF; on-axis FWHM $\sim 6\arcsec$)
at any redshift of interest\footnote{$ 100\arcsec > R_{c} >
20\arcsec$ for $0.1 <z<1$}, it is incorrect to assume that all
clusters brighter than a given flux will be detected \---\ unless
the flux limit is set to some high value. Cluster detectability
depends not only on the instrumental PSF, object flux and
morphology but also upon the background level and the detector
topology (e.g. CCD gaps and vignetting), in addition to the
ability of the pipeline to separate close pairs of point-like
sources \---\ all of which are a function of the specified energy
range \citep{scharf02}. We thus stress that the concept of ``sky
coverage'', i.e. the fraction of the survey area covered at a
given flux limit, is strictly valid only for point-sources
because, for faint extended objects, the detection efficiency is
surface brightness limited (rather than flux limited). Moreover,
since the faint end of the cluster luminosity function is poorly
characterised at $z>0$, it is not possible to estimate  a
posteriori what   fraction of groups   remain undetected, unless a
cosmological model is assumed, along with a thorough modelling of
the cluster population out to high redshift; the lower limit of
the mass or luminosity function being here a key ingredient. To
our knowledge, this has never been performed in a fully
self-consistent way so far for any deep X-ray cluster survey
($``F_{lim}" \sim 2-5 \times 10^{-14}$ \flux ). It is also
important to consider that the flux recorded from a particular
cluster represents only some fraction of the total emitted flux
and must therefore be corrected by integrating an
assumed spatial emission model to large radius.\\
Consequently, with the goal of constructing deep controlled
samples suitable for cosmology   we define, rather than   flux
limits,  classes of extended sources. These are defined in the
extension and significance parameter space and  correspond to
specific levels of contamination and completeness.  As shown by
\citet{pacaud06}, extensive simulations of various cluster and AGN
populations generate detection probabilities as a function of
sources properties. This enables a simultaneous estimate of the
source completeness levels and of the frequency of contamination
by misclassified point-like sources or spurious detections. X-ray
source classification was performed using {\tt Xamin} and employs
the output parameters: {\tt extent, likelihood of extent,
likelihood of detection}. The reliability of the adopted selection
criteria has been checked against the current sample of 60
spectroscopically confirmed XMM-LSS clusters.  We have defined
three classes of extended sources:
\begin{itemize}

\item The C1 class is defined such that no point sources are
misclassified as extended and is described by {\tt extent}~$>
5\arcsec$, {\tt likelihood of extent}~$>33$ and {\tt likelihood of
detection}~$>32$.  The C1 class contains the highest surface
brightness extended sources and inevitably includes a few nearby
galaxies \---\ these are readily discarded from the sample by
inspection of optical overlays.

\item The C2 class is described by {\tt extent}~$>5\arcsec$ and
$15<$~{\tt likelihood of extent}~$<33$ and typically displays a
contamination rate of 50\%. The C2 class includes clusters fainter
than C1, in addition to a number of nearby galaxies. Contaminating
sources include saturated point sources, unresolved pairs, and
sources strongly masked by CCD gaps, for which not enough photons
were available to permit reliable source characterization.
Contaminating sources were removed after a visual inspection of
the optical/NIR data for each field   and in some cases as a
result of follow-up spectroscopy.

\item The C3 class was constructed in order to investigate
clusters at the survey sensitivity limit, particularly clusters at
high redshift. Sources within the C3 class typically display
$2\arcsec<$~{\tt extent}~$<5\arcsec$ and {\tt likelihood of
extent}~$>4$. Selecting such faint, marginally extended sources
generates a high contamination rate. However, low selection
thresholds are required to identify extended sources at the survey
limit: faint sources will never be characterized by high
likelihood values. When refining the C3 class, the X-ray, optical
and NIR appearance was examined thoroughly. Generally speaking, C3
sources display low surface brightness or extended emission
affected by a point source. Additional constraints included that
the detection should be located at an off-axis angle $< 10\arcmin$
and that the total detection should generate 30 photons or greater
(stronger constraint on the off-axis value is necessary because
weak objects are subject to strong distortions beyond 10 arcmin,
thus hardly measurable). The most plausible C3 candidates were
investigated spectroscopically and confirmed clusters are
presented.

\end{itemize}

The analysis of simulated cluster and AGN data permits the
computation of selection probabilities for the C1 and C2 cluster
samples\footnote{The C1 and C2 classes are defined from
simulations representative of a mean exposure time of 10 ks. In
the present paper, we keep the same definition as the
signal-to-noise ratio (S/N) increase is only $\sqrt 2$.}. The
extent to which the C1 and C2 classes are comparable to flux
limited samples is analyzed in detail by \citet{pacaud06} and
further discussed in the last section of this paper. The selection
probability for C3 clusters has not been determined.

\subsection{Determination of cluster redshifts}

 The XMM-LSS spectroscopic Core Programme aims at the redshift
confirmation  of the  X-ray  cluster candidates; velocity
dispersion may subsequently  be obtained for a sub-sample of
confirmed clusters as a second step programme. Spectroscopic
observations were performed using a number of telescope and
instrument combinations and are summarized in Table 2. Details of
which observing configuration was employed for each cluster are
presented in Table 3.

The minimum criterion required to confirm a cluster was specified
to be three concordant  redshifts ($\pm 3000$~kms$^{-1}$) within a
projected scale of approximately 500~kpc of the X-ray emission
centroid, computed at the putative cluster redshift. For nearby
X-ray clusters of temperature $T_X=2$~keV, a radius of 500~kpc
corresponds to approximately 50\%\ of the virial radius and
encloses 66\%\ of the total mass \---\ with both fractions being
larger for higher temperature clusters \citep{arnaud05}. The final
cluster redshift was computed from the non weighted mean of all
galaxies within this projected aperture and within a rest-frame
velocity interval $\pm 3000$~km$s^{-1}$ of the interactively
determined redshift peak. Potential cluster galaxies are selected
for spectroscopic observation by identifying galaxies displaying a
uniformly red colour distribution within a spatial aperture
centred on the extended X-ray source (see Andreon et al. 2004 for
more details). Cluster members flagged via this procedure are then
allocated spectroscopic slits in order of decreasing apparent
magnitude (obviously avoiding slit overlap).

The exact observing conditions for each cluster form a heterogeneous
distribution. However, each cluster was typically observed with a
single spectroscopic mask featuring slitlets of 8--10\arcsec\ in
length and 1--1\farcs4 in width. The use of a different telescope and
instrument configuration generally restricts the available candidate
cluster member sample to a different limiting $R$--band magnitude
(assuming an approximately standard exposure time of 2 hours per
spectroscopic mask). Typical apparent $R$--band magnitude limits
generating spectra of moderate (${\rm S/N} > 5$) quality for each
telescope were found to be the following: VLT/FORS2 (23),
Magellan/LDSS2 (22) and NTT/EMMI (21.5). Spectroscopic data reduction
followed standard IRAF\footnote{IRAF is distributed by the National
Optical Astronomy Observatories, which are operated by the Association
of Universities for Research in Astronomy, Inc., under cooperative
agreement with the National Science Foundation.} procedures. Redshift
determination was performed by cross-correlating reduced,
one-dimensional spectra with suitable templates within the IRAF
procedure {\tt xcsao} \citep{kurtz98} and confirmed via visual
inspection. A more detailed description of the spectroscopic
techniques employed by the XMM-LSS survey can be found in
\citet{val04} and \citet{willis05}. In addition to the above
spectroscopic observations, cluster redshift information was
supplemented where available by spectra contributed from the VVDS (see
Table 3).

\begin{table*}
\caption{Observing resources employed to determine cluster spectroscopic redshifts.}
\label{spectro}
\centering
\begin{tabular}{lllcc}
\hline
Telescope & Instrument  & Grism $+$ Filter & Approximate resolving power ($R$) & Identifier \\
\hline\hline
VLT & FORS2 &  300V$+$GC435  & 500  & 1 \\
VLT & FORS2 & 600RI$+$GC435 & 1000 & 2 \\
VLT & FORS2 & 600z          & 1300 & 3 \\
NTT & EMMI  & Grism \#3     & 700  & 4 \\
Magellan (Clay) & LDSS-2 & medium-red & 500 & 5 \\
VLT & VIMOS & LRRED & 220 & 6 \\
\hline
\end{tabular}
\end{table*}

The D1 X-ray clusters  with confirmed redshifts are presented in
Table 3. C1 and C2 confirmed clusters constitute a controlled
sample (following Sec. 2.3) and  are associated with the label
``XLSSC'' and a three digit identifier\footnote{The acronym is
defined at CDS at the following URL {\tt
 http://vizier.u-strasbg.fr/cgi-bin/Dic?XLSSC}.}. This
nomenclature is used to identify individual clusters in any later
discussion.  The completeness of C3 sources is not addressed. In
the case where a particular cluster is present in two separate XMM
pointings, only the pointing where the cluster is the closest to
the optical centre has been used to measure its properties. Note
that the off-axis restriction imposed on the  C3 clusters excludes
two faint clusters located at the very border of the D1 area and
reported in the \citet{willis05} initial sample
(XLSSUJ022633.9-040348, XLSSUJ022628.2-045948). Cluster redshift
values given in Table 3 are the unweighed mean of relatively small
member samples and observed, in a few cases, using different
spectrographs. As this approach may result in relatively large
(several hundred kms$^{-1}$) uncertainties in the computed
redshift, the cluster redshift precision reported in Table 3 has
been set to two decimal points (3000~km$s^{-1}$).
X-ray/optical overlays of each cluster field are displayed in Fig.
\ref{overlay}.

\begin{table*}
\label{list}
\begin{minipage}{180mm}
\caption{Spectroscopically confirmed X-ray clusters within the D1 area.}
\begin{tabular}{lcccccccc}
\hline
Source   & XLSSC & RA (deg.) & Dec (deg.) &  XMM & Off-axis
angle\footnote{The off-axis angle is computed  from the barycentre
of the optical axes of the three telescopes using XMMSAS variables
XCEN YCEN weighted by the mean detector sensitivity (see
\citet{pacaud06}).}
& Redshift & \#\ of members\footnote{Only  galaxies within a projected distance $<500$~kpc of  the cluster centre are counted.} & Observed \\
              &       &           &            &   pointing   & (arcminutes)   &          &                   & (see Table 2) \\
\hline\hline
C1 && &    & & & & \\ \hline
XLSS J022404.1-041329\footnote{Listed by \citet{andreon05}.}   & 029  & 36.0172 & -4.2247 & G03 & 9.0  & 1.05    & 5   & 3 \\
XLSS J022433.5-041405    & 044  & 36.1410 & -4.2376  & G03 & 3.8  & 0.26    & 9   & 4 \\
XLSS J022524.5-044042    & 025  & 36.3526 & -4.6791  & G07 & 10.3 & 0.26    & 10  & 5 \\
XLSS J022530.6-041419    & 041  & 36.3777 & -4.2388  & G02 & 9.1  & 0.14    & 9   & 4 \\
XLSS J022609.7-045804    & 011  & 36.5403 & -4.9684  & G11 & 8.1  & 0.05    & 7   & 4 \\
XLSS J022709.1-041759\footnote{Already published by \citet{val04}.}   & 005  & 36.7877 & -4.3002     & G01 & 7.8  & 1.05    & 5   & 2 \\
XLSS J022725.8-043213\footnote{Already published by  \citet{willis05}.}    & 013  & 36.8588 & -4.5380  & G05 & 8.1  & 0.31    & 11  & 5 \\
XLSS J022739.9-045129    & 022  & 36.9178 & -4.8586  & G10 & 5.6 &
0.29    & 5   & 5 \\  \hline
 C2 && &    & & & & \\ \hline XLSS
J022725.0-041123   & 038 & 36.8536 & -4.1920  & G01 & 1.9  & 0.58
& 7   & 4 \\ \hline
 C3 && &    & & & & \\ \hline
XLSS J022522.7-042648   & a  & 36.3454 & -4.4468  & G07 & 3.9  & 0.46    & 4   & 2 \\
XLSS J022529.6-042547    & b  & 36.3733 & -4.4297  & G07 & 5.8  & 0.92    & 7   & 6 \\
XLSS J022609.9-043120    & c  & 36.5421 & -4.5226  & G06 & 8.0  & 0.82    & 8   & 6 \\
XLSS J022651.8-040956   & d & 36.7164 & -4.1661  & G01 & 6.6  & 0.34    & 5   & 1 \\
\hline
\end{tabular}
\end{minipage}
\end{table*}

\section{X-ray properties of confirmed clusters}

The spectral and spatial X-ray data for each spectroscopically
confirmed cluster was analysed to determine the temperature, spatial
morphology and total bolometric luminosity of the X-ray emitting gas.

\subsection{Spectral modelling and temperature determination}
\label{spectralfit}

A complete description of the spectral extraction and analysis
procedures as applied to X-ray sources with low signal levels
using the {\tt Xspec} package \citep{karnaud96}, together with a
discussion on the accuracy of the computed temperatures, are
presented in \citet{willis05}. We summarise the principal steps
below.

\begin{figure}
   \centering
   \includegraphics[width=8cm]{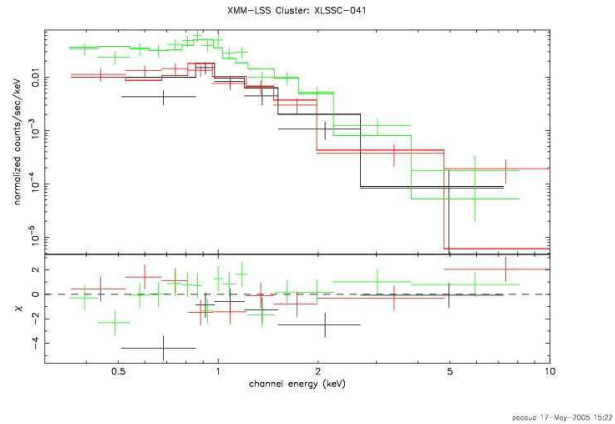}
      \caption{ Fitted spectrum and residuals for cluster XLSSC 041; (MOS1: black, MOS2:
red, pn: green) }
         \label{spec041}
   \end{figure}

Spectral data were extracted within an aperture of specified
radius (see Table \ref{specfit}) and a corresponding background
region was defined by a surrounding annulus.  Photons were
extracted over the energy range [0.3--10]~keV, excluding the
energy range [7.5--8.5]~keV due to emission features produced by
the pn detector support. Analyses of simulated spectral data with
less than 400 total counts indicated that using C-statistics on
unbinned spectral data produced a systematic offset in the
computed temperature. This bias is significantly reduced for such
faint spectra when the data are resampled such that at least 5
photons are present in each spectral bin corresponding to the
background spectrum. We determined that this approach produces
reliable temperature measures for low temperature ($T<3$~keV), low
count level ($<400$ photons) spectral data. The assumed fitting
model employs an absorbed APEC plasma \citep{smith01} with a fixed
metal abundance ratio given by \citet{grevesse99} and set to 0.3
of the solar value. Absorption due to the Galaxy is modelled using
the WABS function \citep{morrison83}, fixing the hydrogen column
density to the value given by \citet{dickey90} at the cluster
position (typically $\sim 2.6 \times 10^{20}$ cm$^{-2} $). Where
the temperature fitting procedure failed to converge to a sensible
model (due to low signal levels), the source temperature was fixed
at 1.5~keV. Results of the  X-ray spectral analysis are presented
in Table \ref{specfit}. An example of cluster spectrum and fit is
shown in Figure \ref{spec041}.

\begin{table*}
\caption{Spectral X-ray parameters. The radius of the circular
aperture used for the spectral extraction is denoted by
$R_{spec}$. Computed source counts are summed over the three
detectors. The suffix ``F'' applied to temperature values
indicates that a reliable temperature fit was not achieved and a
gas temperature of 1.5~keV was assumed for the computation of
bolometric luminosity.} \label{specfit} \centering
\begin{tabular}{cccccccc}
\hline
Cluster & $R_{spec}$ & Source counts in $R_{spec}$ & $T$     &    C-stat   & \multicolumn{2}{c}{$r_{500}$}   \\
        & (arcscec)  &  [0.3-7.5]+[8.5-10] keV & (keV) & (per d.o.f) & (Mpc) & (arcsec)                \\
\hline
\hline
XLSSC 029 & 33 & 311    & $4.1_{-1.0}^{+1.7}$ & 1.08  & 0.52 & 67      \\
XLSSC 044 & 55 & 234    & $1.3_{-0.2}^{+0.2}$ & 1.15  & 0.40 & 100  \\
XLSSC 025 & 35 & 661    & $2.0_{-0.3}^{+0.5}$ & 1.06  & 0.53 & 129 \\
XLSSC 041 & 45 & 523    & $1.3_{-0.1}^{+0.3}$ & 1.00  & 0.43 & 172  \\
XLSSC 011 & 68 & 425    & $0.6_{-0.1}^{+0.2}$ & 1.04  & 0.28 & 272  \\
XLSSC 005 & 35 & 164    & $3.7_{-1.4}^{+3.5}$ & 1.02  & 0.49 & 60   \\
XLSSC 013 & 30 & 161    & $1.0_{-0.2}^{+0.2}$ & 0.92  & 0.33 & 73  \\
XLSSC 022 & 39 & 1304   & $1.7_{-0.2}^{+0.2}$ & 0.91  & 0.47 & 109   \\
\hline
XLSSC 038 & 33 & 118    & 1.5F                & \---\ & 0.37 & 56 \\
\hline
cluster a & 24 & 160    & 1.5F                & \---\ & 0.40 & 69  \\
cluster b & 30 & $<100$ & 1.5F                & \---\ & 0.30 & 38 \\
cluster c  & 30 & $<100$ & 1.5F                & \---\ & 0.32 & 42    \\
cluster d & 25 & 157    & $0.9_{-0.2}^{+0.2}$ & 0.74  & 0.31 & 65  \\
\hline
\end{tabular}
\end{table*}

\subsection{Source morphology and spatial modelling}
\label{spatial_fit}

Sources detected using {\tt Xamin} are initially compared to two
surface brightness models describing the two dimensional photon
distribution: a point source and a circular $\beta$-profile of the form
\begin{equation}
{
S(r) = \frac{A}{[1+(r/R_{c})^{2}]^{3\beta-1/2}},
}
\end{equation}
where $\beta = 2/3$ is fixed while the core radius, $R_{c}$, and
profile normalisation, $A$, are permitted to vary
\citep{cavaliere76}. Each profile is convolved with the mean
analytical PSF at the corresponding off-axis location and a
comparison of the statistical merit achieved by each profile
provides an effective discriminator of point and extended sources
in addition to an initial estimate of the source flux (see
\citet{pacaud06}).

The photometric reliability of this procedure when applied to
faint, extended sources is affected by the presence both of gaps
in the XMM CCD array and by nearby sources (although both are
described within the fitting procedure) \---\ largely due to
variations in the true source morphology and the fact that a
larger fraction of the total emission is masked by the background
when compared to brighter sources.  Although such effects are
naturally incorporated into the selection function appropriate for
each cluster class via simulations \citep{pacaud06}, a further
interactive spatial analysis was performed on each
spectroscopically confirmed cluster in order to optimize the
measure of the total emission (i.e. flux and luminosity) within a
specified physical scale.

The photon distribution for each confirmed cluster is modelled
using a one-dimensional circular $\beta$-profile in which $\beta$,
$R_{c}$ and $A$ are permitted to vary. Photons from the three XMM
detectors are co-added applying a weight derived from the relevant
exposure map and pixels associated with nearby sources are
excluded. Photons are binned in concentric annuli of width
3\arcsec\ centred on the cluster X-ray emission. Radial data bins
are subsequently  resampled to generate a minimum $S/N>3$  per
interval. The background is computed at large radius assuming a
constant particle contribution plus vignetted cosmic emission. The
above resampling procedure is then applied to the circular
$\beta$-profile convolved with the mean analytical
PSF\footnote{The convolution of the two profiles models the photon
distribution factor introduced by the two dimensional convolution
\citep{arnaud02}.}  computed at the corresponding cluster off-axis
angle \citep{ghizzardi02}. Model cluster profiles are realised in
this manner over a discrete grid of $\beta$ and $R_c$ values with
the best-fitting model for each cluster determined by minimising
the $\chi^{2}$ statistic over the parameter grid.  Finally, the
best-fitting spatial profile (at this point in units of photon
count rate) is integrated out to a specified physical radius and
converted into flux and luminosity units using standard procedures
within {\tt Xspec}.

The majority of confirmed clusters are apparently faint,
displaying total photon counts of order a few hundred, and the
observed photon distribution in many cases represents only a
fraction of the extended X-ray surface brightness distribution.
Under such conditions the parameters $\beta$ and $R_{c}$ are
degenerate when fitted simultaneously, limiting the extent to
which ``best-fitting'' parameters can be viewed as a physically
realistic measure of the cluster properties, although providing a
useful ad hoc parametrisation. For this reason we do not quote
best fitting values of $\beta$ and $R_{c}$ derived for each
confirmed cluster. The uncertainty associated with the procedure
is evaluated using a large suite of simulated observations (and
subsequent analyses) of clusters of specified surface brightness
properties (i.e. $\beta$, $R_{c}$ and apparent brightness - Pacaud
et al., in preparation). The fractional uncertainty  can then be
quoted as a function of the number of photons collected within the
fitting radius $R_{fit}$ (the maximum radius out to which the
resampling criterion ${\rm S/N} > 3$  was achieved) and the radius
to which the profile is   calculated (possibly, extrapolated).
Note that, as shown in the next section, almost all clusters have
$R_{fit}$ greater than the physical integration radius, hence
requiring no profile extrapolation. For the very faintest clusters
(those with total photon counts less than $\sim$ 100) a simple sum
within a circular aperture was applied.

\begin{figure}
\centering
\includegraphics[width=8cm]{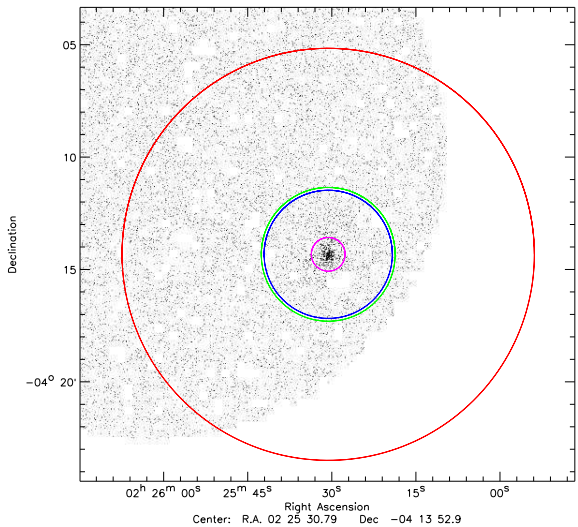}
\caption{An example of the spatial and spectral analysis regions
applied to cluster XLSSC 041. The photon image is displayed. The
purple circle indicates the spectral extraction region. The green
circle indicates $r_{500}$. The blue circle indicates $R_{3\sigma}
= R_{fit}$. All X-ray sources, except the cluster of interest, are
masked. The external red circle delineates the region used for the
fit.} \label{spa_mask}
\end{figure}

\begin{figure}
\centering
\includegraphics[width=8cm]{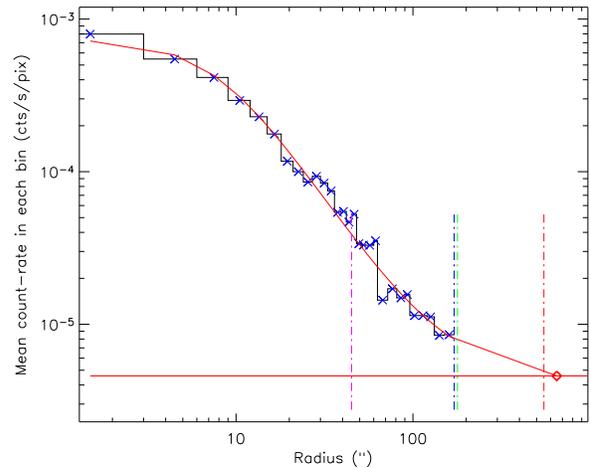}
\caption{The radially averaged  X-ray emission profile for cluster
XLSSC 041. The red curve indicates the best-fitting model compared
to the data (black histogram with blue crosses). The vertical
lines follow the same colour coding as Figure \ref{spa_mask}. The
red horizontal line marks the background level.} \label{spa_prof}
\end{figure}

As an example of the above spatial fitting procedure, Figure
\ref{spa_mask} shows the data analysis regions applied to the cluster
XLSSC 041. Figure \ref{spa_prof} displays the resulting one
dimensional radial profile and the best-fitting surface brightness
model for the same cluster.

\subsection{Determination of cluster flux and luminosity}
\label{fluxlum}

Values of flux and luminosity for confirmed clusters are obtained by
integrating the cluster emission model, described by the appropriate
{\tt Xspec} plasma emission and surface brightness models, out to a
specified physical radius. We use a different physical radius for flux
measures as opposed to luminosity - mainly because tabulated flux
values for cluster surveys present in the literature prefer an
estimate of the ``total'' flux within a limited energy interval
whereas luminosity values are computed as the bolometric emission
within a physical radius corresponding to a constant overdensity in an
evolving universe (e.g. $r_{500}$).

Flux values are computed by integrating the best-fitting
$\beta$-profile to a radius of 500 kpc. The specified aperture
includes a substantial fraction of the total flux (approximately
2/3 of the flux from $\beta$-profile described by $\beta=2/3$ and
$R_{c} = 180$~Mpc) yet avoids uncertainties associated with the
extrapolation of the profile to large radii.

In order to obtain cluster luminosities within a uniform physical
radius, we have integrated the best--fitting $\beta$-profile for
each cluster to $r_{500}$ i.e. the radius at which the cluster
mass density reaches 500 times the critical density of the
universe at the cluster redshift.  Values of $r_{500}$ for each
cluster were computed using the mass-temperature data of
\citet{fino01} which, when converted to our assumed cosmology and
fitted with an orthogonal regression line, yield the expression
$r_{500} = 0.391~T^{0.63}~h_{70}(z)^{-1}$~Mpc for clusters
in  the  $0.7<T<14$~keV interval.  As reported in Willis et al.
(2005), values of $r_{500}$ from this formula agree well with
those derived from assuming an isothermal $\beta$-profile for
clusters displaying $T \lesssim 4$~keV.

For each cluster in the confirmed sample, with the exception of
XLSSC 005, the computed values of $r_{500}$ lie within radius of,
or close to,  the region employed to fit the $\beta$-profile;
(hence,   for this cluster alone, the photometric errors are large
as reported in Table \ref{spacefit}). Bolometric X--ray
luminosities were calculated for each cluster by extrapolating the
APEC plasma code corresponding to the best--fitting temperature to
an energy of 50~keV. Values of $r_{500}$, flux and luminosity for
each confirmed cluster are listed in Tables \ref{specfit} and
\ref{spacefit}.

In Appendix \ref{appA} we analyse the impact of further sources of
uncertainty affecting the luminosity and temperature measurements.

Appendix \ref{appB} gather notes on the individual clusters and
investigate, among others, the possibility that the computed
clusters brightness values may be contaminated by AGN emission. We
conclude that none of the C1 and C2 clusters (i.e. those used for
detailed population statistics) contained in the D1 sample are
significantly contaminated by AGN emission.

\begin{table*}
\caption{Results of the spatial analysis for confirmed clusters.
See text for the definition of the fitting radius $R_{fit}$. The
net source counts are computed within $R_{fit}$ and are
uncorrected for vignetting. A value of ``NF'' indicates that no
reliable spatial fit was possible for the cluster:  the source
counts, flux and luminosity were computed applying a circular
aperture of radius 500 kpc.
\newline
Flux values are computed by integrating the best-fitting cluster
$\beta$-profile out to a radius of 500 kpc.  The photometric
precision indicates the mean $1~\sigma$ errors estimated from
analyses of simulated cluster data and accounts for the profile
fitting uncertainties only (see text for details). }

 \label{spacefit} \centering

\begin{tabular}{l  c c c c c   }
\hline\hline
Cluster & $R_{fit}$  & Source counts in $R_{fit}$   & $F_{[0.5-2]} $   & $L_{bol}(r_{500})$ &  Photom. acc.  \\
&  arcsec & in [0.5-2] keV &   $10^{-14}$ erg/s/cm$^{2}$  & $10^{44}$ erg/s &  \\

 \hline
  XLSSC 029   & 60 & 361  & 3.1 &4.8 &20\% \\
  XLSSC 044 &129 &318  &  2.5&0.11&15\%   \\
  XLSSC 025 &123 & 905& 9.4&0.52 &15\%  \\
  XLSSC  041 & 171 &819 & 20.6& 0.24 &15\%  \\
  XLSSC 011 & 354  & 972 &  16.4 & 0.015 &15\%   \\
  XLSSC 005 &39  & 128 &1.1 & 1.5 &60\%    \\
  XLSSC 013& 234   & 383 & 2.7& 0.15 &20\%   \\
  XLSSC 022 & 171  & 1785 &  9.8 &0.65  &10\%      \\ \hline
XLSSC 038& NF &[60] & 0.3 & 0.09 & - \\ \hline
cluster a & NF & [108]&  0.7&0.1   & -\\
cluster b & NF & [52]  & 0.4& 0.4   &-   \\
cluster c & NF  & [29] &   0.3 & 0.3  & - \\
cluster d&   432   & 417 &  0.83 & 0.078 &20\% \\
\hline
\end{tabular}
\end{table*}

\subsection{Trends in the $L_X$ versus $T_X$ correlation}

\begin{figure}
\centering
\includegraphics[width=8cm]{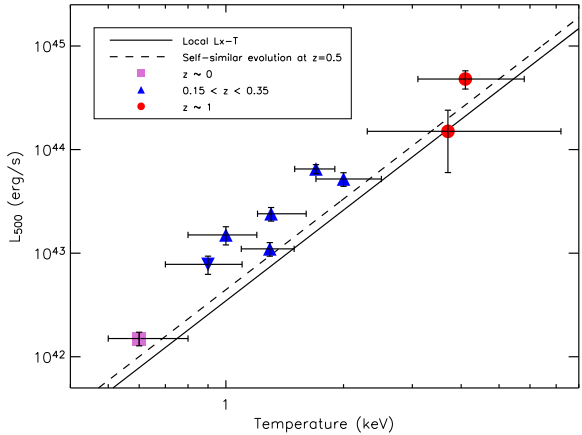}
\caption{$L_X(r_{500})$ versus $T_X$ relation for the clusters for
which it was possible to derive a temperature; all of them but
{\em cluster d} (displayed as an upside-down triangle) are C1. The
solid line gives the mean local $L_X-T_X$ relation (see text),
while the dotted line is the expected luminosity enhancement
assuming self-similar luminosity scaling within $r_{500}$ at
$z=0.5$. Different plotting symbols indicate clusters located
within three different redshift intervals.} \label{lumt}
\end{figure}

Figure \ref{lumt} displays the $L_X$ versus $T_X$ distribution of
the D1 clusters for which it was possible to measure a temperature
(eight C1 and one C3 objects).  Although the D1 area represents
only a subset of the anticipated XMM-LSS area, the C1 sample is
complete and reliable temperature information is available for all
systems. It is therefore instructive to consider trends in the
$L_X$ versus $T_X$ distribution in anticipation of a larger sample
of C1 clusters from the continuing survey. The location of C1
clusters in the $L_X$ versus $T_X$ plane is compared to a
regression line computed for a combined sample of local sources
based upon the group data of \citet{osmond04} and cluster data of
\citet{markevitch98}. The computed regression line takes the form,
$\log L_X = 2.91~ \log T_X +42.54$, for bolometric luminosity
computed within $r_{500}$. A complete discussion of the regression
fit will be presented by \citet{helsdon06}.

One issue of interest concerns the properties of intermediate
redshift ($z\sim 0.3$) X--ray groups (i.e. $T_X \sim 1-2$~keV).
Such systems dominate the XMM-LSS numerically and, when compared
to higher temperature, higher mass clusters, are expected to
demonstrate to a greater degree the effects of non--gravitational
physics in the evolution of their X--ray scaling relations with
respect to self--similar evolution models. The luminosity of X-ray
sources in XMM-LSS may be compared to those of local sources at
the same temperature by computing the luminosity enhancement
factor, $F=L_{obs}/L_{pred}$, where $L_{obs}$ is the observed
cluster X--ray luminosity within a radius, $r_{500}$, and
$L_{pred}$ is the luminosity expected applying the fitted $L_X$
versus $T_X$ relation computed for the local sources and the
XMM--LSS measured temperature. Sources XLSSC 013, 022 and 041 have
a luminosity enhancement factors $F \approx 3-4$, compared to a
value 1.15 expected from self--similar\footnote{Self-similar
implies that the luminosity scales as the Hubble constant when
integrated within a radius corresponding to a fixed ratio with
respect to the critical density of the universe as a function of
redshift \citep{voit05}.} luminosity scaling within $r_{500}$.

From the local universe, we know that low temperature groups show
a larger dispersion in the L-T relation than massive clusters
\citep{helsdon03}. This reflects their individual formation
histories, since they are particularly affected by
non-gravitational effects, as well as the possible contributions
from their member galaxies. The apparent biasing toward more
luminous objects and/or cooler system could come from the fact
that we detect more easily objects having a central cusp, i.e.
putative cool-core groups. This has the effect of both decreasing
the temperature and increasing the luminosity. The bias could also
simply reflect the fact that we can measure a temperature only for
the brightest objects.

In order to test this hypothesis, we have considered a few orders
of magnitude. The local L-T relation predicts a luminosity of
$1.1~ 10^{43}$ and $2.6~ 10^{43}$ erg/s for a T=1.5 and T=2 keV
group respectively. A factor of 2 under luminosity for such
objects would thus correspond to $5.6~ 10^{42}$ and $1.3 ~
10^{43}$ erg/s. In Table \ref{spacefit}, we note that (i)  the
lowest flux cluster (XLSSC 005) is detected with some 150 photons
in $R_{fit}$; the X-ray image appears moreover to be quite flat;
(ii) group XLSSC 044 (z= 0.26) has a luminosity of $1.1 ~ 10^{43}$
erg/s, for some 300 photons in $R_{fit}$. From this, we infer that
we could have detected groups around $z \sim 0.25$, having $1.5<T<
2 $ keV  that are under luminous by a factor of 2, if any were
present in our sample. Below 1.5 keV, these objects are likely to
remain undetected.

The coming availability of the larger XMM-LSS sample will permit a
more reliable assessment of such effects on the
morphology--luminosity--temperature plane for such groups (Pacaud
et al, in preparation).

\section{Discussion and conclusion}

We have used 20~ks XMM images to construct a deep sample of galaxy
clusters. The total cluster surface density of 15.5~deg$^{-2}$ is approximately five times larger than achieved
previously with the deepest ROSAT cluster surveys (e.g. RDCS,
\citet{borgani01}). On the one hand, from the optical point of
view, we note that none of the detected clusters shows strong
lensing features, hence the likely absence of massive clusters in
the D1 area\footnote{With the caveat that the presence of giant
arcs requires not only a large mass concentration but also a
specific lens/source configuration}. This is consistent with the
fact that the highest cluster X-ray temperature is only 4 keV:
this temperature corresponds to a mass of $\sim
3~10^{14}~M_{\sun}$ and, in a standard $\Lambda$CDM halo model
\citep{pacaud06}, the density of clusters more massive than this
limit, i.e. those most likely to produce strong lensing, is $\sim 0.8$~deg$^{-2}$. On the other hand, it is indeed a salient
property of the XMM-LSS to unveil for the first time the bulk of
the $1<T< 2$ keV group population in the $0.1<z<0.4$ range along
with its capability of detecting $z\geq1$ clusters. We further
review below the main properties of the sample.

The C1 cluster sub-sample corresponds to a purely X-ray selected
sample (zero contamination) and displays a surface density of $\sim 9$~deg$^{-2}$. Reliable temperature information is available
for all C1 sources and optical spectroscopic observations are
required only to confirm the cluster redshift. Relaxing the
selection criteria used to generate the C1 sample creates
additional samples labelled C2 and C3. However, optical imaging
and spectroscopic data are required to identify bona-fide clusters
within these samples. The C2 sample possesses a well-defined X-ray
selection function (approximately 50\%\ of sources are confirmed
as clusters) and the surface density of C1+C2 clusters is
$\sim 11$~deg$^{-2}$. Sources labelled C3 represent significant
detections outwith the C1 and C2 selection criteria and, given the
high contamination rate, we do not compute a selection function
for these sources. The C3 sample  contains potentially interesting
objects and points to our ultimate sensitivity for cluster
detection which appears to be around $5\times10^{-15}$ \flux;
however, we stress that the full selection function for the C1 and
C2 samples is multi-dimensional (see below).  Noting this caveat
\---\ for comparison purposes only \---\ the quoted flux
sensitivity corresponds to a cluster of $\sim 7\times10^{43}$ erg
s$^{-1}$ ($T \sim 3$~keV) at $z=1$ and to a group of
$3.5\times10^{42}$ erg s$^{-1}$ ($T=1$~keV) at $z=0.3$. We further
note that no C1 or C2 cluster emission appears to be significantly
contaminated by AGN activity \---\ partly a result of the high
threshold put on the extent likelihood for these samples.

Having the D1 XMM-LSS sample now assembled, it is instructive to
examine in what manner it differs from a purely flux limited
sample. This question is phenomenologically addressed  by
\citet{pacaud06} as the answer depends on two major ingredients;
(1) the pipeline efficiency (involving itself the many
instrumental effects) - this is quantified by means of extensive
simulations;  (2) the characteristics of the cluster population
out to a redshift of unity at least - this latter point being
especially delicate as the low-end of the cluster mass function,
critical for the survey sensitivity, is poorly known and cluster
scaling law evolution, still a matter of debate. Hence the need
for a self-consistent approach basically involving a cosmological
model, a halo mass spectrum and some $L_{X}(M,~T,~R_{c},~z)$
function\footnote{The function  is normalized from local universe
observations and its evolution constrained by available high-z
data, numerical simulations and other possible prescriptions such
as self-similarity evolution; one of the main unknowns being the
role of non gravitational physics in cluster evolution} describing
the evolution of the cluster intrinsic properties that directly
impact on the cluster detection efficiency. The principal
conclusion regarding the use of a single flux limit is that, to
obtain a cluster sample displaying a high level of completeness
and reasonably low contamination, the flux limit has to be set to
some high value, e.g. $F > 5 \times 10^{-14}$ \flux\ in the case
of XMM-LSS. The present study demonstrates  on real data the
advantage of using the C1 set of criteria  as a well defined
sample that includes groups down to $T$ = 1 keV and fluxes as low
as $10^{-14}$ \flux\ and, consequently, significantly increases
the size of the purely X-ray selected sample. Although the C1
criteria, even fully controlled, might at first sight appear more
pipeline dependent and, thus, less physical than a simple flux
limit, we stress that any X-ray detection algorithm is bound to
miss low luminosity clusters in a way that is pipeline dependent -
the loss of efficiency not being a simple flux limit. For these
reasons, we favor the concept of {\em controlled sample} (in the
C1 selection sense) rather than of {\em complete flux limited
sample}.

Our D1 sample  contains 7 clusters displaying a flux in excess of
$2 \times 10^{-14}$~\flux\ (all C1). This corresponds to a surface
density of $\sim 8$~deg$^{-2}$ and is larger than the 4--5
clusters deg$^{-2}$ implied by the RDCS $\rm \log N - \log S$
relation \citep{rosati98} \---\ the only ROSAT cluster sample
complete to $2 \times 10^{-14}$ \flux\ \citep{rosati02} \---\ in
addition to the shallow XMM/2dF survey \citep{gaga05}, reporting
$\rm 7/2.3 = 3 \, deg^{-2}$ at the same flux limit. The
probability to obtain 8 clusters  $\deg^{-2}$ from the RDCS number
density is 5-10 \%, assuming simple Poisson statistics. Given the
relatively small fields covered in each case, the effects of
cosmic variance upon any such comparison may well be important
(the XMM/2dF survey and deepest regions of the RDCS cover 2.3 and
5~\dd\ respectively). For comparison, the simple cluster evolution
plus cosmological model presented by \citet{pacaud06} predicts a
surface density of $\sim7.5$ clusters deg$^{-2}$ displaying
$T>1$~keV and a flux $F > 2 \times 10^{-14}$ \flux.

We have computed reliable temperature values for nine of the
thirteen confirmed clusters \---\ in particular, the C1 sample is
``temperature'' complete. This is important as it displays the
potential for survey quality XMM observations to investigate the
evolution of X-ray (and additional waveband) cluster scaling
relations in a statistical manner over a wide, uniformly-selected
redshift interval. Together with the initial sample presented by
\citet{willis05}, the D1 sample is the first sample to investigate
the $L_X$ versus $T_X$ relation for $1<T<2$~keV groups at
$0.15<z<0.35$, for the simple reason that this population was
previously undetected. XMM-LSS therefore samples a relatively
complete, high surface density population of clusters displaying
temperatures $T>1$~keV at redshifts $z \lesssim 1$ and provides an
important new perspective for the study the cluster and group
evolution employing only moderate XMM exposure times.

All data presented in this paper \---\ cluster images taken at X-ray
and optical wavebands in addition to detailed results for the spectral
and spatial analyses \---\ are available in electronic form at the
XMM-LSS cluster online database: {\tt
http://l3sdb.in2p3.fr:8080/l3sdb/login.jsp}.

\section*{Acknowledgments}

We are grateful to M. Arnaud for providing her profile convolution
routine.  We thank J. Ballet, R. Gastaud and J.-L. Sauvageot for
useful discussions. A. Gueguen acknowledges a CNES CDD position.
G. Galaz and H. Quintana thank the support of the FONDAP Center
for Astrophysics \# 15010003. S. Andreon   acknowledges financial
contribution from contract ASI-INAF I/023/05/0. The results
presented here are based on observations obtained with XMM-Newton,
an ESA science mission with instruments and contributions directly
funded by ESA Member States and NASA. The cluster optical images
were obtained with MegaPrime/MegaCam, a joint project of CFHT and
CEA/DAPNIA, at the Canada-France-Hawaii Telescope (CFHT) which is
operated by the National Research Council (NRC) of Canada, the
Institut National des Sciences de l'Univers of the Centre National
de la Recherche Scientifique (CNRS) of France, and the University
of Hawaii. This work is based in part on data products produced at
TERAPIX and at the Canadian Astronomy Data Centre as part of the
Canada-France-Hawaii Telescope Legacy Survey, a collaborative
project of NRC and CNRS.

\bibliography{mnrasmnemonic,XMM-LSS_D1_c.bbl}
\bibliographystyle{mnras}

\pagebreak

\appendix

\section{Error budget on the cluster luminosity and temperature
measurements}

 \label{appA}

\begin{table*}
\caption{Error budget for secondary effects.
\newline
$r_{500}^{-}$ and $r_{500}^{+}$ are the extreme possible values
for $r_{500}$ derived from the temperature uncertainties quoted in
Table \ref{specfit}}
 \label{error} \centering
\begin{tabular}{l c c c }
\hline
XLSSC & 41 &   29 & 13    \\
\hline\hline source counts in $R_{fit}, ~R_{spec}$ & 819, 523 &
361, 311 &  383, 161\\
 \hline
$r_{500}^{-},~r_{500},~r_{500}^{+}$  & 171, 179, 196 arcsec& 66, 55, 82 arcsec& 76, 69, 84 arcsec\\
$\frac{L_{X}(r_{500}^{-/+})-L_{X}(r_{500})}{L_{X}(r_{500})}$ &
-2\% /+ 4\%
&-1\% /+ 0.6\% & -5\% /+ 6\% \\
 \hline
Ab = 0.1 & T = 1.3 (1.1, 1.5) keV & T = 4.2 (3.1, 6.1)
keV & T = 1.0 (0.8, 1.2) keV \\
Ab = 0.3 & T = 1.3 (1.2, 1.6) keV & T = 4.1 (3.0, 5.8)
keV & T = 1.0 (0.9, 1.2) keV\\
Ab = 0.6 & T = 1.5 (1.3, 1.7) keV & T = 4.0 (3.0, 5.5)
keV & T = 1.0 (0.9, 1.3) keV \\
 \hline
Undetected AGN contribution & & & \\
$R<$ 500 kpc & 0.6 \% & 0.5 \% & 0.8 \% \\
 \hline
\end{tabular}
\end{table*}

This Appendix investigates the impact of specific sources of
uncertainty on the cluster temperature and flux measurements. The
calculations are performed for 3 clusters representative of the
flux and temperature ranges covered by the present sample, namely
XLSSC 41, 29, 13.

-1- Cluster luminosities are computed within $r_{500}$, a quantity
empirically derived from the temperature (Sec. \ref{fluxlum}),
while the photometric errors quoted in Table \ref{spacefit}
results from the spatial fit only. In Table \ref{error} we provide
the uncertainty on the luminosity  induced by the errors from the
temperature measurements,  as propagated through the derivation of
$r_{500}$ (the spatial fit is then assumed to be exact). The
results show that they are negligible compared to the accuracy
level of the spatial fits.

-2- Cluster temperatures are estimated fixing the metal abundance
to 0.3 solar (Sec. \ref{spectralfit}). Since many of our clusters
have $T< $2 keV, for which the contribution from emission lines is
significant,   some temperature-abundance degeneracy could occur
in the spectral fitting - all the more so since the number of
photons involved in the spectral fit is small. In Table
\ref{error} we provide further temperature measurements fixing the
abundance to 0.1 or 0.6 solar. The results show that the impact on
the derivation of the temperature and associated errors is
negligible. In all these trials, the Galactic column density is
held fixed.

-3- Finally, we investigate whether the contribution of unresolved
AGNs to the  integrated cluster emission is statistically
significant at our sensitivity. We proceed here assuming the XMM
LogN-LogS for point sources \citep{moretti03} since no information
on the AGN environment of low-luminosity clusters is available to
date.  For each ring of the cluster spatial profile, we compute
the limiting flux for which a point source is to be detected at
the 3 $\sigma$ level. We then integrate the LogN-LogS over the
cluster area out to $R = 500$ kpc, between the varying flux limit
and the background flux level. Results are gathered in table
\ref{error} and show that, statistically, the point source
contribution is negligible. In Appendix \ref{appB} we further
inspect, for each cluster, the possibility that the core of the
cluster emission could be contaminated by an  AGN.

\vspace{10cm}

\section{Notes on individual clusters}
\label{appB}

In this Appendix, we provide additional information for confirmed
clusters. We have paid particular attention to whether the X-ray
spectra and spatial emission of each cluster displays evidence for
contamination by AGN emission. For each  cluster we have compared
the results generated by fitting the spatial emission having
either included or excluded the central radial bin (radius of 3 or
6  arcsec.). Although in some cases the best-fitting values of
$\beta$ and $R_{c}$ varied significantly, in all cases the
integrated countrate within $r_{500}$ showed variations less than
10\%. The small number of photons contributed by the central few
arcseconds of each cluster prevented a separate spectral analysis
of the central regions. Finally, we have investigated all examples
of spatial coincidence between radio sources presented in
\citet{bondi03} and the D1 cluster sample. We discuss individual
clusters below.

In parallel, we have discovered a point-like source, XLSS
J022528.2-041536, which is associated, at least in projection,
with a group at redshift $z\sim0.55$ (more than 5 concordant
redshifts). However, the group emission, which is embedded in that
of XLSSC 041, appears totally dominated by  the point source and
consequently, this object does not enter the C1 or C2 or C3
classes. A radio source ($S_{tot} = 0.28$~mJy, $\sim 5\arcsec$
extent) is present at 2\arcsec\ from the X-ray centre. XLSS
J022528.2-041536  is located in the field of XLSSC 041 and the
group member galaxies are indicated  in the overlay of Fig. 6. In
a further study involving optical, radio and IR data, we will
assess the fraction of clusters that remain unclassified  because
of strong AGN contamination.

\begin{description}

\item[{\bf XLSSC 029:}] The presence of a radio source ($S_{tot} =
1.5$~mJy, $\sim 2\arcsec$ extent) some 11\arcsec\ from the cluster
centre might suggest that the X-ray emission is contaminated by an
AGN.  However, examination of a recent 100 ks XMM pointing on this
object (obs 0210490101, PI L. Jones) shows that the astrometry of
the original XMM-LSS survey image is correct within 1\arcsec, and
does not reveal a secondary maximum coincident with the radio
source.

\item[{\bf XLSSC 044:}] This cluster is of very low surface
brightness and displays elongated emission. Its X-ray morphology,
galaxy distribution and temperature of $\sim 1$~keV suggest a
group in formation.

\item[{\bf XLSSC 025:}] A weak radio source ($S_{tot} = 1.0$~mJy, $\sim
5\arcsec$ extent) lies at the centre of the cluster emission, however,
the central galaxy spectrum displays no strong emission lines
(rest-frame wavelength interval $3000-6000\AA$ is sampled). The X-ray
spatial profile is peaked yet exclusion of the central 3\arcsec\
radial bin, does not change significantly the fitted value of $R_c$
(it changes from 5\arcsec\ to 6\arcsec\ while $\beta= 0.44$ remains
constant). We thus exclude any strong contamination by a central AGN,
and favor the cool core hypothesis.

\item[{\bf XLSSC 041:}] No radio source was identified within
30\arcsec\ of the cluster emission centroid. But,

\item[{\bf XLSSC 011:}] The X-ray and optical appearance of this
system are suggestive of a compact group of galaxies. All point
sources within a radius of~ 350\arcsec\ have been removed from the
spatial analysis. However, it was not possible to estimate the
extent of any contribution from the central galaxy to the group
emission.

\item[{\bf XLSSC 005:}] No radio sources are identified within the
projected area covered by the X-ray emission. This system displays
double peaked X-ray emission morphology, with each peak associated
with a distinct velocity component. The photometric uncertainty
for this system is large as less than 200 photons were available
for the spatial fit which was limited to $R_{fit}\sim r_{500}/2 $.
The system is further discussed in \citet{val04}.

\item[{\bf XLSSC 013:}] No radio sources are identified within
1\arcmin\ of the X-ray centroid. A bright X-ray point source is
present at 1\arcmin\ distance yet does not affect the cluster
classification procedure and the point source is removed from the
subsequent spatial and spectral analysis.

\item[{\bf XLSSC 022:}] The X-ray profile is peaked and here is a weak
radio source within 2\arcsec\ of the X-ray emission centroid ($S_{tot}
= 0.15$~mJy, no extent). The spectrum of the central galaxy displays
no significant emission features consistent with AGN activity. The
cluster morphology is very similar to XLSSC 025 in that $R_c$ and
$\beta$ values do not vary significantly depending upon the inclusion
of the central bin in the spatial analysis. We therefore favour the
cool core hypothesis for this system.

\item[{\bf cluster a:}] The X-ray centroid coincides with a radio
source ($S_{tot}$ = 0.21  mJy, no extent) within 2 arcsec. The
optical spectrum obtained for this object is faint and did not
allow us to secure its redshift; however, no emission line is
apparent in the spectrum. The photometric redshift is 0.98, with a
SBI starburst as best fitting spectrum. The flux at 24 $\mu$m is
0.4 mJy which is rather high. This suggests that the coincidence
between this red object and the X-ray centroid might be
fortuitous, but the contamination by an active nucleus cannot be
excluded.

\item[{\bf cluster b:}] An X-ray point source is located some
30\arcsec\ from the X-ray emission centroid and was subsequently
removed from the spatial and spectral analyses.

\item[{\bf cluster d:}] A radio source is located within 3\arcsec\
of the X-ray emission centroid ($S_{tot} = 0.08$~mJy, no extent).
The optical spectrum of the central galaxy does not show
significant emission features. With an extent of 2\farcs5 and a
{\tt extent likelihood}~$ = 21$ this marginal source is classified
as C3.

\end{description}

\end{document}